\documentstyle[twocolumn,prl,aps,epsfig,amssymb,rotate]{revtex}

\newcommand{\beq}{\begin{equation}}
\newcommand{\eeq}{\end{equation}}
\newcommand{\ba}{\begin{array}}
\newcommand{\bea}{\begin{eqnarray}}
\newcommand{\ea}{\end{array}}
\newcommand{\eea}{\end{eqnarray}}

\newcommand\comment[1]{ \hbox{[{\it Comment suppressed here.}\/]} }
\newcommand\hide[1]{}

\newcommand{\skipover}[1]{}

\begin{document}                                                
\title{Higgs description of two-flavor QCD vacuum}
\author{J{\"u}rgen Berges and Christof Wetterich}
\bigskip
\address{
Institut f{\"u}r Theoretische Physik,
Philosophenweg 16, 69120 Heidelberg, Germany}
\maketitle
\begin{abstract} 
A Higgs description for the QCD vacuum in the limit of two
quark flavors is proposed. Complete ``spontaneous 
breaking'' of color symmetry is
realized by condensation of an adjoint 
quark-antiquark and a quark-quark pair.
The vacuum is invariant under 
isospin symmetry applying color and flavor 
transformations simultaneously. All elementary excitations 
acquire integral charges.
The dressed microscopic quark and gluon degrees of freedom 
can be identified with the macroscopic baryon and vector meson
degrees of freedom. 
In addition to the nucleons we find
soliton--type fermions with integer isospin. 
Their presence provides an 
important test accessible to lattice QCD simulations.

\end{abstract}
\begin{narrowtext}

\section{Introduction}

An understanding of the low energy properties in the theory of 
strong interactions described by quantum chromodynamics (QCD) may be based 
on the complementarity between a Higgs-- and a confinement description of 
the vacuum \cite{CC}. Complementarity has been exploited already for the
high temperature and high density phases of gauge theories.
In the electroweak standard model the Higgs phase at low temperature 
and the high temperature confined phase were shown to be continuously 
connected \cite{EW}. In QCD it has been pointed out that the high 
baryon number density phase admits a Higgs description 
by the phenomenon of color-flavor locking \cite{CFL,CONT,JB}. 
Recently, it has been proposed that the vacuum of QCD
can be characterized by a color octet quark-antiquark condensate
involving three quark flavors \cite{GM}. As a consequence color is 
broken completely and a successful phenomenology 
emerges \cite{GM}. A dynamical mechanism for the formation 
of the octet condensate is provided by the anomalous multiquark 
interaction induced by instantons \cite{CWINST}.

In this Letter we propose a Higgs description for the vacuum of
QCD in the limit of two quark flavors. The Higgs picture gives
a simple relation between the microscopic (quark--gluon)
and the macroscopic (hadron) degrees of freedom, 
which overcomes the classic qualitative problems of an analytic approach to
low-energy QCD.  Both spontaneous chiral 
symmetry breaking and confinement are associated to the condensation
of an adjoint quark-antiquark and an antitriplet quark-quark pair.  
The condensates leave a global vector--like $SU(2)_{{\rm color}+V}$ subgroup
invariant which applies color and flavor transformations simultaneously.
This symmetry is physical isospin. In the presence of both condensates 
all elementary excitations acquire integral charges.  
Quarks can be identified with baryons and
gluons with vector mesons. In contrast to \cite{GM} 
diquark condensation yields a modified conserved baryon number. 
As a consequence the dressed quarks 
acquire integer baryon number, two of them carrying the quantum numbers 
of the proton and the neutron. The quark degrees of freedom and the 
baryon degrees of freedom can thus be described by the same fields. 
Since color is broken completely all gluons acquire a mass by the 
Higgs phenomenon and four of them carry the quantum numbers of the 
$\rho$-- and $\omega$--mesons. The  
pseudo-Goldstone bosons associated to the spontaneous breaking 
of chiral symmetry can be identified with the pions.

In addition to the expected nucleons and mesons 
the remaining quarks and gluons correspond to
soliton-type fermionic excitations which carry zero baryon number  
and bosonic excitations with baryon number one. Excitations with similar 
quantum numbers are known to occur in two-flavor QCD at sufficiently high 
baryon density \cite{pert,ARW}.
In the vacuum these excitations comprise half--integer isospin 
representations of 
bosons and integer isospin representations of fermions \cite{BWW}. 
In the nonrelativistic
quark model for two-flavor QCD these states could only be constructed from an 
infinite number of quarks and antiquarks. 
We emphasize that within our approach the 
appearance of stable nonlinear excitations follows only from 
group-theoretical arguments independent of the detailed dynamics.
This provides an important test for the idea of color symmetry breaking
in the vacuum. Even though two-flavor QCD is an idealization and not 
accessible to direct experimental observation it can be simulated 
by lattice QCD.    

Consequences of a possible
condensation of an adjoint quark-antiquark and a quark-quark pair 
have been recently discussed at high baryon number density \cite{JB}. 
It was found that simultaneous condensation can be observed in the high
density phase for sufficiently strong interactions in the 
respective channels. In this case vacuum and nuclear 
matter in the non-superfluid phase may be continuously connected.

\section{Color-isospin locking}

{\it Color octet, antitriplet and singlet condensates.\ }
In the proposed Higgs description the vacuum can 
be characterized by the condensation of a color-octet quark-antiquark 
pair
\beq
\chi \sim 
\Big\langle \bar{\psi}^{i}_a \sum_{s=1}^3
\left( \tau_s \right)_{ab} \left( \lambda_s \right)^{j i} 
\psi^{j}_b \Big\rangle  
\label{chicond}
\eeq
and a color-antitriplet diquark
\beq
\Delta \sim \Big\langle \psi^{i}_a 
\left( \tau_2 \right)_{ab} \left( \lambda_2 \right)^{ji } 
\psi^{j}_b \Big\rangle
\label{delcond}
\eeq
where $(\tau_s)_{a,b}$ ($a,b=1,2$) are the Pauli matrices in flavor 
space, and $(\lambda_s)^{i j}$ ($i,j=1,2,3$) denote
the Gell-Mann matrices in color space. Here we have suppressed the Dirac 
structure.  

Since the adjoint quark--antiquark condensate 
$\chi$ belongs to a flavor triplet 
it breaks both color and flavor symmetries separately. It is 
invariant under the global vector--like $SU(2)_{{\rm color}+V}$ subgroup
which applies color and flavor transformations simultaneously.
The condensate locks the flavor generators
$\tau_s$ to the transposed color generators $\lambda_s$ of a 
$SU(2)_{\rm color}$--subgroup and extends the notion of 
color-flavor locking to a two-flavor theory.  
Condensation of the form (\ref{chicond}) leaves a local $U(1)_8$ 
subgroup of color unbroken. It gives a mass to seven of the 
eight gluons by the Higgs mechanism.  

The color-antitriplet diquark condensate 
$\Delta$ given in (\ref{delcond}) is a flavor singlet 
and does not break chiral symmetry (see also \cite{pert,ARW}). 
Condensation of this form 
leaves an $SU(2)$ subgroup of $SU(3)_{\rm color}$ 
unbroken. Consequently three gluons remain massless if only
$\Delta$ condenses.

Neither the adjoint chiral condensate nor the antitriplet
diquark condensate alone break color completely. 
Only both condensates together give a mass to all gluons.
Chiral symmetry is spontaneously broken by $\chi \not = 0$ and four quarks 
get a mass even in the chiral limit of a vanishing current quark mass 
$m_{\rm q}$. The remaining two quarks become massive for a
nonvanishing singlet condensate 
\beq 
\phi \sim \Big\langle \bar{\psi}^{i}_a  \psi^{i}_a \Big\rangle  \, .
\label{phicond}
\eeq
Further condensates consistent with the symmetries 
include \cite{BWW} a flavor quark-antiquark singlet bilinear 
involving the color $\lambda_8$ generator, or a
pure glue condensate of the form 
$\langle F^{\mu\nu}_{ik} F_{kj \mu \nu} 
- \frac{1}{3} F^{\mu \nu}_{lk} F_{kl \mu \nu} \delta_{ij} \rangle \sim 
\left(\lambda_8\right)_{ij}$. These condensates affect some of the 
quantitative 
aspects but do not change the picture as far as symmetries
are concerned.

\section{Quark-gluon description of hadrons}

{\it Integer charges and modified baryon number.} 
The electromagnetic gauge symmetry with fractional quark charges is broken
in presence of the condensates 
(\ref{chicond}) or (\ref{delcond}).
The vacuum is invariant under a combination 
of quark--electric charge $Q_{\rm q}$ and abelian
color charge $Q_c=\frac{1}{2}\lambda_3+\frac{1}{2 \sqrt{3}} \lambda_8$
of the quarks. As a consequence, the physical electric charge contains 
a color component
\beq
Q=Q_{\rm q} - \frac{1}{2}\lambda_3-\frac{1}{2 \sqrt{3}} \lambda_8  
\eeq
where $Q_{\rm q}=2/3$ for up and $Q_{\rm q}=-1/3$ for down quarks.
One observes that all fields carry integral electric 
charge $Q$. 

The diquark condensate (\ref{delcond}) breaks baryon number.
However, there is a conserved modified baryon
number 
\beq
B'=B_{\rm q} - \frac{\lambda_8}{\sqrt{3}}
\label{newB}
\eeq
with $B_{\rm q}=1/3$ for both quarks. The condensates $\Delta$ and $\chi$
are neutral under $B'$. Electric 
charge is related to $B'$ and the third component of isospin, $I_3$,
by $Q= I_3 + B'/2$.

{\it Quarks.}
We show the quark-electric charge $Q_{\rm q}$, abelian color charge
$Q_c$ and the quantum numbers $Q, I_3$ and $B'$ of the 
six quarks in table 1. There are only two degrees of freedom, 
corresponding 
to the quarks of the third color, which carry baryon number $B'=1$. 
The dressed up quark of the third color carries electric charge 
$Q=+1$ and the corresponding down quark is electrically neutral.
They carry the quantum numbers of the proton and the neutron, respectively.
In presence of the condensates (\ref{chicond}), (\ref{delcond})
and (\ref{phicond}) the nucleons acquire a mass
\beq
M_p^2 = M_n^2 = (m_{\rm q}+\phi)^2 
\eeq
independent of $\chi$ and $\Delta$. The mass matrix for the 
remaining four quark fields contains Majorana--type entries
similar to the case of neutrinos. Indeed, no quantum number 
forbids a mixing of the quark fields with their antiparticles.  
There is an isospin triplet $(S^0,S^+,S^-)$ with charge $Q=(0,1,-1)$ and 
a neutral singlet $(L^0)$ which get contributions from all three condensates  
\bea
M_S^2 &=& (m_{\rm q}+\phi-\chi)^2+\Delta^2 \,\, ,\nonumber\\[0.2ex]
M_L^2 &=& (m_{\rm q}+\phi+3\chi)^2+\Delta^2  \, .
\label{sbaryons}
\eea
\begin{center}
\begin{tabular}{|c|cc|rrc|l|}
\hline
Quarks & $Q_{\rm q}$& $Q_c$ & $Q$ & $I_3\,$ & $\,\,B'\,$ & \\
\hline\hline
$u_3$&$2/3$&$\,-1/3\,$&$1$&$1/2$&$1$&$\, p$\\
$d_3$&$-1/3\,\,\,\, $&$-1/3$&$0$&$-1/2$&$1$&$\, n$\\
\hline
$\,u_1-d_2\,$&$2/3,-1/3$&$\,\,\,\, 2/3,-1/3$&$0$&$0$&$0$&$\, S^0$\\
$u_2$&$2/3$&$-1/3$&$1$&$1$&$0$&$\, S^+$\\
$d_1$&$-1/3\,\,\,\, $&$\,\,\,\,\, 2/3$&$-1$&$-1$&$0$&$\, S^-$\\
\hline
$u_1+d_2$&$2/3,-1/3$&$\,\,\,\, 2/3,-1/3$&$0$&$0$&$0$&$\, L^0$\\
\hline
\end{tabular}\\

\medskip Table 1: Charges of the up and down quarks.
\end{center}

{\it Vector mesons.}
All eight gluons carry integer electric charge. Three gluons
carry the quantum numbers of the
isospin triplet $\rho$--mesons. The isospin singlet corresponding
to the eighth gluon can be associated with the neutral
$\omega$--meson. The adjoint chiral condensate (\ref{chicond}) 
contributes to the $\rho$-meson mass while the diquark condensate
(\ref{delcond}) is responsible for a nonvanishing mass of the  
$\omega$--meson 
\beq
M_{\rho}^2 = c_{\chi} g^2 \chi^2 \, , \qquad
M_{\omega}^2 = c_{\Delta} g^2 \Delta^2 \, .  
\eeq
The proportionality factors $c_{\chi}$ and $c_{\Delta}$
reflect the relative strength of the coupling of the condensates
to quarks and gluons. The vector meson masses are 
proportional to the strong gauge coupling $g$.
The mass of the two
isospin doublets $(\kappa^+,\kappa^0)$ and 
$(\kappa^-,\bar{\kappa}^0)$ 
get contributions from both colored condensates,
\bea 
M_{\kappa}=\frac{3}{8} c_{\chi} g^2 \chi^2 
+ \frac{3}{4} c_{\Delta} g^2 \Delta^2  \, .
\label{smesons}
\eea 

\medskip
\begin{center}
\begin{tabular}{|c|r|rrr|l|}
\hline
Gluons &$\, Q_c \,$&$Q$&$I_3$&$B'\,\,$&\\
\hline\hline
$A_3$&$0\,$&$0$&$0$&$0\,\,$&$\,\,\rho^0$\\
$\, A_1+iA_2\, $&$\, -1\,$&$1$&$1$&$0\,\,$&$\,\,\rho^+$\\
$A_1-iA_2$&$1\,$&$-1$&$-1$&$0\,\,$&$\,\,\rho^-$\\
\hline
$A_8$&$0\,$&$0$&$0$&$0\,\,$&$\,\,\omega$\\
\hline
$A_4+iA_5$&$-1\,$&$1$&$1/2$&$1\,\,$&$\,\,\kappa^{+}$\\
$A_6+iA_7$&$0\,$&$0$&$-1/2$&$1\,\,$&$\,\,\kappa^0$\\
\hline
$A_4-iA_5$&$1\,$&$-1$&$-1/2$&$-1\,\,$&$\,\,\kappa^{-}$\\
$A_6-iA_7$&$0\,$&$0$&$1/2$&$-1\,\,$&$\,\,\bar{\kappa}^0$\\
\hline
\end{tabular}\\

\medskip Table 2: Charges of the gluons.
\end{center}

{\it Pseudo-Goldstone bosons.}
There is also a triplet of collective modes associated 
with chiral symmetry breaking by the $\chi$ or $\phi$ condensates. 
The broken symmetry generators
are given by the axial charges and the bosons with mass squared
$m^2_{\pi} \sim m_{\rm q}$ match the quantum numbers of the three pions.    
The interactions of the pions are dictated by spontaneously broken
chiral symmetry. No Goldstone bosons are generated by $\Delta \not = 0$.

\section{``Strange'' particles}
\label{strange}

{\it Soliton-type excitations.}
As a consequence of the complete breaking of color we
find, apart from the expected nucleons and mesons, additional 
excitations $S,L^0$ and $\kappa$ with unusual properties. These states carry 
integral charges and form part of the spectrum. They can be expected 
to be relatively heavy since their masses 
get contributions from all condensates $\chi,\Delta$ and $\phi$ (cf.\
eqs.\ (\ref{sbaryons}), (\ref{smesons})). 
From table 1 one observes that there are integer
isospin representations of fermions, a triplet $(S,S^+,S^-)$
and a singlet $(L)$, which carry zero baryon number. Table 2 shows
half-integer isospin representations of bosons,
two doublets $(\kappa^+,\kappa^0)$ and $(\kappa^-,\bar{\kappa}^0)$,
with baryon number one.
Their presence could be tested in lattice Monte Carlo simulations
of two-flavor QCD. Such a test is not completely straightforward 
since integer isospin fermions and half-integer isospin bosons 
do not couple in the channels of standard operators representing
a finite number of quarks. Nontrivial operators in the gluon
sector are needed.

For an understanding of this it is instructive to consider
the nonrelativistic quark model. In the quark model for
two-flavor QCD the building blocks of hadrons are up and down
quarks belonging to isospin doublets. No half--integer  
isospin representations can occur as $\bar{q}q$--states or any bosonic state
involving a finite number of quarks $q$ and antiquarks $\bar{q}$.
In fact, there is a simple selection rule that the product
of a finite even number of half--integer isospin representations
has integer isospin and an odd number has half--integer isospin.
Bosons build from a finite number of fermions must involve an
even number and therefore carry integer isospin. We conclude
that the bosonic $\kappa$--states can arise in the two-flavor 
quark model only as soliton-type excitations involving infinitely
many quarks and antiquarks. Similarly, the fermions with
integral isospin ($S,S^+,S^-$ and $L$) cannot be represented by a 
finite odd number of quarks or antiquarks in the quark model.   

The $\kappa$--states carry integer $B'$. They can
therefore not decay into $\rho$--mesons and pions (or photons) 
which all carry $B'=0$. The decay into protons and neutrons involves an 
even number of fermions with $B'=\pm 1$. It would therefore lead to
a final state with $B'$ even and is again 
forbidden by $B'$ conservation. The $\kappa$--states are stable if
their mass is below the masses of $S$ and $L^0$. 
The (lightest of the) latter ones are stable
since they carry $B'=0$ and cannot decay into protons and
neutrons. Fermion number is only conserved modulo two, as is manifest
already from the Majorana-type entries in the mass matrix for
the $S$-- and $L$--states.

{\it ``Strangeness''.}
The quantum numbers of the additional
particles become more familiar if seen from a three-flavor
perspective. In terms of standard three-flavor hadronic quantum numbers
one may write $B'=$ ``$B$''$+$ ``$S$'' for ``baryon number $B$'' and 
``strangeness $S$''.
One observes from table 1 that with 
the association ``$B$''$=1$, ``$S$''$=-1$ for the triplet
of $S$--states there is a precise 
correspondence to the quantum numbers of the hyperon 
$\Sigma$-triplet. Similarly,
$L^0$ can be associated with the singlet $\Lambda^0$.  From table 2
one finds that the $\kappa$-states with ``$B$''$=0$ can be 
mapped to the $K^*$-mesons. In this language the appearance
of ``strange'' particles is a consequence of the fact that
color is broken and the definition of ``$S$'' contains a color
component. Despite the apparent correspondence
we emphasize
that there is no additional $U(1)$ symmetry associated to strangeness
present in the two-flavor theory.

\section{High density or temperature}

\mbox{\it High density and Higgs description of nuclear matter.} 
The 
appearance of ``strange'' states in the proposed vacuum of two-flavor QCD
follows only from the symmetry breaking pattern and is independent of 
the detailed dynamics. It is interesting to note that  
fermionic states carrying zero baryon number 
are known from two-flavor QCD at high baryon number density 
where diquark condensation of the form (\ref{delcond}) 
occurs \cite{pert,ARW}. The high
density ground state is neutral under $B'$ given
in eq.\ (\ref{newB}) and exhibits four fermionic excitations
with $B'=0$. A major difference to the vacuum is that
at asymptotically high density no adjoint chiral condensate is 
present and, in particular, all fermionic excitations 
carry half--integer isospin.
 
At nuclear matter densities, however, a nonvanishing adjoint chiral condensate
cannot be excluded. A Higgs description of nuclear matter 
with simultaneous condensation of quark-quark 
and adjoint quark-antiquark pairs can be achieved at moderate densities 
if sufficiently strong interactions in the respective channels are present 
\cite{JB}. This opens the possibility that nuclear matter 
in the non-superfluid phase can be continuously connected to
the vacuum. At zero temperature nuclear matter is expected to be 
in a superfluid phase due to nucleon pairing breaking isospin.  
In the Higgs description the quarks of the 
third color carry the same quantum numbers as the proton and the 
neutron. The possible channels for condensation correspond precisely 
to the known possibilities for pairing in nuclear matter 
(see also \cite{CONT}). If isospin is broken there is one exactly 
massless boson associated to superfluidity with the quantum numbers of 
the $a_0$--meson. We note that two light scalars correspond to the 
breaking of the generators $I^\pm$. The masses of the corresponding
$a^+$-- and $a^-$--mesons vanish in the limit of equal up and down 
quark masses.

{\it High temperature.\ }  
Lattice simulations for two-flavor QCD at high temperature 
and zero density find that chiral symmetry restoration and 
deconfinement occur approximately at the same temperature
$T_c$ \cite{Lattice}. Sufficiently below $T_c$ the equation of state
is rather well approximated by a gas of hadrons. Above $T_c$ the 
dominant thermodynamic degrees of freedom are gluons and quarks.
In the present Higgs description both spontaneous chiral symmetry  
breaking and confinement are associated to the condensation
of quark-antiquark and quark-quark pairs. The transition to the 
high temperature state proceeds by the melting of these condensates.

Qualitatively different aspects of the transition are described 
by the differing condensates.
The adjoint chiral condensate breaks both chiral symmetry and color up to 
an abelian $U(1)$ gauge symmetry. If $\chi$ becomes zero at some critical
temperature $T_c$ then chiral symmetry is restored {\it and} at the same 
temperature part of the gluons become massless. This provides a simple 
mechanism for the connection between chiral symmetry restoration and 
deconfinement in QCD at high temperature. 
 
For $\Delta = 0$ all quarks
carry fractional baryon number $B_q = 1/3$. 
A vanishing diquark condensate signals the hadron-quark transition.
For $\chi \not = 0$ the absence of the diquark condensate has a comparably
small effect on the gluon sector. Only the $\omega$-meson becomes massless.

Whether or not the hadron--quark transition happens at the temperature 
when all gluons become massless, i.e.\ if $\chi$ and $\Delta$ 
vanish at the same $T_c$, is a question of the detailed dynamics.
Also the singlet chiral condensate $\phi$ will play a role for the 
quantitative aspects of chiral symmetry restoration.\\    

Two flavor QCD is an idealization which is expected to ressemble in 
many aspects realistic QCD. The proposed Higgs description, in particular,
predicts intriguing features as the existence of soliton-type
excitations. Lattice Monte Carlo simulations can constitute important tests
of this picture.\\

We thank Mark Alford and Uwe-Jens Wiese for very helpful discussions.

\end{narrowtext}


\begin{references}

\bibitem{CC} T. Banks, E. Rabinovici, Nucl.\ Phys.\ {\bf B160} (1979)
349;
E.\ Fradkin, S.\ Shenker, Phys.\ Rev.\ {\bf D19} (1979) 3682;
G.\ 't Hooft, in: {\it Recent Developments in Gauge Theories},
(Plenum, New York, 1980) p.\ 135;
S.\ Dimopoulos, S.\ Raby, L.\ Susskind, Nucl.\ Phys.\ {\bf B173}
(1980) 208;
T.\ Matsumoto, Phys.\ Lett.\ {\bf 97B} (1980) 131;
M. Yasu\`e, Phys.\ Rev.\ {\bf D42} (1990) 3169.

\bibitem{EW} M.\ Reuter, C.\ Wetterich, Nucl.\ Phys.\ {\bf B408}
(1993) 91; C.\ Wetterich, ``Electroweak physics and the Early Universe'', eds.\
J.\ Romao, F.\ Freire, Plenum Press (1994) 229; W.\ Buchm{\"u}ller,
O.\ Philipsen, Nucl.\ Phys.\ {\bf B443} (1995) 47; K.\ Kajantie,
M.\ Laine, R.\ Rummukainen, M.\ Shaposhnikov, Phys.\ Rev.\ Lett.\
{\bf 77} (1996) 2887.

\bibitem{CFL}
M. Alford, K. Rajagopal, F. Wilczek, Nucl. Phys. {\bf B537} (1999) 443;

\bibitem{CONT}
T. Sch\"afer, F. Wilczek, Phys.\ Rev.\ Lett.\ {\bf 82} (1999) 3956;
M. Alford, J. Berges, K. Rajagopal, Nucl.\ Phys.\ {\bf B558} (1999) 219;
T. Sch\"afer, F. Wilczek, Phys.\ Rev.\ {\bf D60} (1999) 074014. 

\bibitem{JB} J.\ Berges, hep-ph/0012013.

\bibitem{GM} C.\ Wetterich, Phys.\ Lett.\ {\bf B462} (1999) 64; 
hep-ph/0008150; hep-ph/9908514, appears in Eur.Phys.J.\ {\bf C}.
\bibitem{CWINST}
C.\ Wetterich, hep-ph/0011076.

\bibitem{pert}
D.\ Bailin, A.\ Love, Phys.\ Rep.\ {\bf 107} (1984) 325; D.T.\ Son, 
Phys.\ Rev.\ {\bf D59} (1999) 094019;
D.H.\ Rischke, D.T.\ Son, M.A.\ Stephanov, hep-ph/0011379. 

\bibitem{ARW}
M.\ Alford, K.\ Rajagopal, F.\ Wilczek, Phys.\ Lett.\ {\bf B422} (1998) 247;
R.\ Rapp, T.\ Sch\"afer,
E.V.\ Shuryak, M.\ Velkovsky, Phys.\ Rev.\ Lett.\ {\bf 81} (1998) 53.

\bibitem{BWW}
J.\ Berges, U.-J.\ Wiese, C.\ Wetterich, unpublished.

\bibitem{Lattice} See e.g.\ F.\ Karsch and E.\ Laermann, A.\ Peikert,
hep-lat/0012023 and references therein.



\end{references}
\end{document}